\documentclass[11pt,letterpaper]{JHEP3}
\usepackage{epsfig}
\usepackage{amssymb}

\newcommand{\beq}{\begin{equation}}
\newcommand{\eeq}{\end{equation}}
\newcommand{\beqa}{\begin{eqnarray}}
\newcommand{\eeqa}{\end{eqnarray}}
\newcommand{\be}{\begin{equation}}
\newcommand{\ee}{\end{equation}}
\newcommand{\bea}{\begin{eqnarray}}
\newcommand{\eea}{\end{eqnarray}}
\newcommand{\bA}{\begin{array}}
\newcommand{\eA}{\end{array}}
\newcommand{\bc}{\begin{center}}
\newcommand{\ec}{\end{center}}

\newcommand{\drawsquare}[2]{\hbox{%
\rule{#2pt}{#1pt}\hskip-#2pt
\rule{#1pt}{#2pt}\hskip-#1pt
\rule[#1pt]{#1pt}{#2pt}}\rule[#1pt]{#2pt}{#2pt}\hskip-#2pt
\rule{#2pt}{#1pt}}

\newcommand{\Yfund}{\raisebox{-.5pt}{\drawsquare{6.5}{0.4}}}

\usepackage{epsfig}
\usepackage{psfrag}
\usepackage{subfigure}
\usepackage{cite}
\usepackage{epstopdf}

\def\be{\begin{equation}}
\def\ee{\end{equation}}
\def\bea{\begin{eqnarray}}
\def\eea{\end{eqnarray}}

\title{Stable Non-Supersymmetric Throats in String Theory}

\author{Shamit Kachru$^{1,2,3}$, ~{\normalsize \bfseries \sffamily Du\v  san Simi\' c$^{1,2,3}$ and Sandip P. Trivedi$^{4,1,2}$}

~\\${}^1$Stanford Institute for Theoretical Physics, Stanford University\\
Stanford, CA 94305 USA \\ \vspace{0.3cm}

${}^2$ SLAC, Stanford University\\
Stanford, CA 94309 USA \\ \vspace{0.3cm}

${}^3$ Kavli Institute for Theoretical Physics\\
Santa Barbara, CA 93106 USA \\ \vspace{0.3cm}

${}^4$ Tata Institute for Fundamental Research\\
Mumbai, 400005, India \\ \vspace{0.3cm}

}

\newcommand{\bbea}{\begin{equation} \begin{aligned}} \newcommand{\eeea}{\end{aligned} \end{equation}}

\abstract{
We construct a large class of  non-supersymmetric AdS-like throat geometries in string theory by  taking non-supersymmetric orbifolds of supersymmetric backgrounds.  The scale of SUSY breaking is the
AdS radius, and the dual field theory has explicitly broken supersymmetry. The large hierarchy of energy scales in these geometries is stable. We establish this by showing that the dual gauge theories
do not have any relevant operators which are singlets under the global symmetries. When the geometries are embedded in a compact internal space, a large enough discrete subgroup of the global symmetries can still survive to prevent any singlet relevant operators from arising. We  illustrate this by
embedding one case  in a non-supersymmetric orbifold of a Calabi-Yau manifold.
These examples can serve as a starting point for obtaining Randall-Sundrum models in string theory, and more generally for constructing composite Higgs or technicolor-like models where strongly coupled dynamics leads to the breaking of electro-weak symmetry.
Towards the end of the paper, we briefly discuss how bulk gauge fields can be incorporated  by introducing D7-branes in the bulk, and also show how the strongly coupled dynamics can lead to an emergent   weakly coupled gauge theory in the IR with matter fields including scalars.

}

\preprint{NSF-KITP-09-55, SITP-09/17, SLAC-PUB-13593}

\def\be{\begin{equation}}
\def\ee{\end{equation}}
\def\bea{\begin{eqnarray}}
\def\eea{\end{eqnarray}}

\begin{document}

\section{Introduction}
The Randall-Sundrum model \cite{RS,RStwo} provides a solution to the hierarchy problem different from supersymmetry.  The essential idea is to have a five dimensional $AdS_5$-like warped spacetime which can give rise to a large hierarchy of scales. By suitably locating
the standard model fields in such a spacetime and tying the hierarchy of scales in the warped background to the breaking of electroweak symmetry, one can try to  construct a workable model of electroweak symmetry breaking.

In this paper we will take some steps towards constructing such a model in string theory.
String compactifications which realize some of the basic physics of the Randall-Sundrum model
have already been described in \cite{HV,GKP} and many subsequent papers.
However, these constructions are based on compactification of SUSY-preserving AdS throat
geometries.  Therefore, the first question we must address is: can a geometry with a long
throat, and thus a large hierarchy of scales, be stable in the absence of SUSY?

The worry is that once SUSY is broken, relevant operators can be generated in the dual field
theory (say at the cutoff, where one glues the throat into a compact geometry).  These would
destroy the throat geometry.  These operators, in the gravity picture, correspond to modes
that grow rapidly in the interior of the throat (as one goes towards the infra-red), and cause
the throat to pinch-off at a high scale.  In this paper, we argue that one can construct
non-supersymmetric throat
geometries (and compactifications incorporating them) that avoid this particular problem.

The basic idea is to find theories with sufficiently rich global symmetries to forbid all
relevant operators.  In theories with scalars, one operator which cannot be forbidden by linearly realized symmetries
is a singlet scalar mass of the form $\phi^\dagger \phi$; but as observed in \cite{Strassler},
such operators obtain large anomalous dimensions in the limit of strong 't Hooft coupling,
and are dual to string states.  Combining these ingredients, we obtain non-supersymmetric
theories with only
marginal or irrelevant global singlet perturbations.  Any global singlet marginal perturbations, if they become marginally
relevant, can end the throat geometry, but only after a long period of RG flow, leaving  a
macroscopic throat.  Our goal will be to realize this picture in concrete examples, much in
the spirit of \cite{Strassler}.\footnote{
For clarity, we should describe what has been learned on top of the basic picture advocated in
\cite{Strassler}.  We believe the present work demonstrates that the existence of examples is much
more generic than one might have believed, that one can suitably ``round off'' such examples
in the IR in a way that allows generation of composite scalars, and that one can compactify
these throats while preserving sufficient discrete symmetries to protect the hierarchy.}

One might wonder whether such constructions, if they exist, would be very contrived or
non-generic.  We will actually find, on the contrary, that the simplest gauge/gravity
dual pairs give rise rather easily to large classes of examples.
We start with the famous duality between ${\cal N}=4$ super Yang-Mills and type IIB
string theory on $AdS_5 \times
S^5$ \cite{Juan}.  We find that an infinite class of non-supersymmetric orbifolds of this dual
pair \cite{orbCFT,LNV}
preserve $SU(3) \times U(1)$ global symmetries, and have all the properties required
to realize our scenario.  In this class, we do not yet have a concrete handle on how the
throat may round off in the deep IR (when marginally relevant operators in the UV have
grown strong).

To obtain a model where we have a slightly more complete picture of the IR physics, we then
turn to the theory
of D3-branes and D5-branes in the conifold geometry \cite{KW,KS}.  We show that simple
orbifolds of this theory again break SUSY while maintaining the absence of global singlet
relevant operators.   In this case, we can use the Klebanov-Strassler solution \cite{KS} to give a
picture of the IR physics which should govern some examples.  We further exhibit a concrete
compactification of such a SUSY-breaking throat, preserving sufficient global
symmetries to stabilize the hierarchy.

Given a concrete model with the IR geometry of the throat under control, we can discuss in more
detail how one might obtain gauge and matter fields at low energies.  We show that two promising
avenues are to realize bulk gauge fields via D7-branes stretching down the throat, or to have
an emergent gauge sector arise in the IR on anti-D3 branes localized at the tip.  The latter
system also gives rise to light composite Higgs-like scalars, which break the gauge symmetry
at low energies.

The full organization of our paper is as follows.
In \S2, we present an infinite class of non-supersymmetric orbifolds of $AdS_5 \times S^5$ which have no
global singlet relevant operators.  We then turn, in \S3, to
brief overview of the Klebanov-Witten (KW) and Klebanov-Strassler (KS) theories of D-branes at
the conifold.  \S4 details the construction of non-supersymmetric orbifolds of these theories
which have the required properties.
In \S5, we turn to the construction of compactifications which incorporate such throats, and
show that the full compact geometry can maintain a large enough subgroup of the global
symmetry group to still forbid all relevant operators.   Having established the existence of
such stable throat geometries, we turn in \S6 to adding (very crude) toy models
that give rise to interesting IR physics (i.e., some weakly interacting sector with light
gauge bosons, fermions, and scalars).
In appendix A, we provide a detailed discussion of how a discrete group we use in \S5\ acts on operators
of the KS field theory, while in appendix B, we discuss some issues related to
the Horowitz-Orgera-Polchinski instability of non-supersymmetric orbifolds \cite{joegary}.

\section{Orbifolds of ${\cal N}=4$ Theory}

We begin by considering a large class of non-supersymmetric orbifolds of the ${\cal N}=4$ theory. These   have a dual description as non-supersymmetric orbifolds of IIB string theory on $AdS_5\times S^5$
\cite{orbCFT,LNV}. The ${\cal N}=4$ theory has a global  $SO(6)$ R  symmetry group, which corresponds in the gravity description to  the isometries of the $S^5$. In the examples we consider, after orbifolding, this global symmetry is broken to an  $SU(3)\times U(1)$ subgroup. We show that at large 't Hooft coupling, where the supergravity description is valid, there are no relevant operators in the gauge theory which are singlets under the surviving global symmetries. This shows that the throat geometry is stable in all  these examples even though  supersymmetry is broken. \footnote{Note that the supersymmetry breaking is not soft, and occurs in the ultraviolet. At large $N$, some properties of the daughter theory are however inherited from the parent SUSY theory.}

Before proceeding, we should describe the current state of knowledge about non-supersymmetric
orbifolds of AdS/CFT \cite{Adams,DKRone,DKRtwo}.  The non-supersymmetric orbifolds with fixed points of the orbifold action
on $S^5$ have twisted sector closed-string tachyons in the gravity regime.  The instability represented
by these tachyons seems to correlate with weak-coupling Coleman-Weinberg instabilities in
the dual field theories (at small 't Hooft coupling) \cite{Adams}.

On the other hand, freely-acting orbifolds (which are the only kind we consider) do not generate any apparent tachyonic instabilities in the gravity regime -- the twisted sector strings have large positive mass proportional to the size of the space.
Therefore, at strong 't Hooft coupling, the AdS/CFT correspondence suggests that there is a large-$N$
fixed line, destabilized by $1/N$ corrections.

In the small radius regime of freely-acting orbifolds, the story is different.  The field theory at
weak 't Hooft coupling and large-$N$ is not at a fixed point (even at the planar level), due to
the generation of double-trace couplings \cite{DKRone,DKRtwo}.

A more subtle issue in these examples is possible non-perturbative instabilities at strong
't Hooft coupling. Horowitz, Orgera and Polchinski \cite{joegary}
analyzed non-perturbative decay channels of precisely the orbifolds we consider.
We summarize their analysis, and its implications for our constructions, in appendix B.

\subsection{Constructing the field theories}

The metric of $AdS_5\times S^5$ spacetime is,
\be
\label{metricads}
ds^2={r^2\over L_{\rm Ads}^2} (-dt^2+dx_i^2)+{L_{\rm AdS}^2\over r^2} dr^2 + L_{\rm AdS}^2 ~d\Omega_5^2.
\ee
Here $d\Omega_5^2$ is the volume element of a unit $S^5$, and $L_{\rm AdS}$ is the common radius of the $AdS_5$ and the $S^5$.
This geometry is obtained as the near-horizon geometry of $D3$-branes which extend along, $t,x_i, i=1,\cdots 3,$ and are transverse to the six coordinates $y^1, \cdots y^6$ \cite{Juan}.

The orbifold we consider is obtained by identifying configurations in IIB string theory related by the action of the ${\mathbb Z}_{k}$ generator:
\be
\label{genzn}
\alpha=R_{({2\pi\over k})}(-1)^F.
\ee
Here,
\be
\label{defrot}
R_{({2\pi \over k})} =\exp[{2\pi i\over k}(J_{12}+J_{34}+J_{56})],
\ee
 is a simultaneous rotation by the angle ${2\pi\over k}$
in the $y^1-y^2, y^3-y^4,$ and $y^5-y^6,$ planes. This rotation acts on the   $S^5$ and leaves the $AdS_5$ invariant.
The second factor in eq.(\ref{genzn}),  $(-1)^F$, weights spacetime fermions and bosons with opposite signs. We only consider the case where  $k$ is an  odd integer.  In this case,  due to the presence of  the $(-1)^F$ factor, $\alpha^k$ is unity on both spacetime fermions and bosons.\footnote{If $k$ is even the orbifold would project out all spacetime fermions and correspond to an orbifold of Type 0 string theory, as described in e.g., \cite{joegary}. We do not consider this case here.}

The orbifold has no fixed points. If $Z^1=y^1+iy^2, Z^2=y^3+iy^4, Z^4=y^5+iy^6$ are the three complex coordinates transverse to the $D3$ branes, then under the action of $\alpha$,
 \be
 \label{actalz}
 (Z^1,Z^2,Z^3)\rightarrow (\exp[{{2\pi i \over k}}]Z^1, \exp[{{2\pi i \over k}}] Z^2, \exp[{{2\pi i \over k}}] Z^3).
 \ee
 The only fixed point would be  at $Z^1=Z^2=Z^3=0$ but this point is not present in the near horizon geometry, where the flux blows up the $S^5$ to non-zero  radius.
 This makes it relatively easy to determine the spectrum of light states. In the supergravity approximation, these states are  simply  those KK modes of the  $AdS_5\times S^5$ background which  are invariant under the orbifold symmetry. The masses of these  modes (and the dimensions of the dual operators) are the same as in the ${\cal N}=4$ theory.

To understand what subgroup of the $SO(6)$ R-symmetry group is preserved by the orbifold, let us note that the $S^5$
   can be described as a $U(1)$ fibration over $CP^2$.
 The orbifold is obtained by identifying points   along the fiber circle related by a shift of $2\pi \over k$.
 This preserves the $SU(3)$ symmetry of the base $CP^2$ and also the $U(1)$ symmetry corresponding to continuous shifts along the fiber.
Thus the $SO(6)$ global symmetry of the ${\cal N}=4$ theory is broken to an $SU(3)\times U(1)$ subgroup after the orbifold identification.

To preserve supersymmetry the  orbifold must have $SU(3)$ holonomy.
$SO(6)\sim SU(4)$ has  a $4$ dimensional spinor representation
$(\psi^1,\psi^2,\psi^3,\psi^4)$, where the first three components transform as a triplet of  $SU(3)$ and the last is a singlet. Under the orbifold symmetry, $\alpha$,
\be
\label{transformsp}
(\psi^1,\psi^2,\psi^3,\psi^4)\rightarrow (-e^{-i\pi\over k} \psi^1, -e^{-i\pi\over k} \psi^2, -e^{-i\pi\over k} \psi^3, -e^{3i \pi \over k} \psi^4)
\ee
(these charge assignments will become clear when we consider the gauge theory below).
For $k=3$ we see that this  leaves $\psi^4$ invariant, so that the resulting holonomy lies in $SU(3)$ and the orbifold preserves ${\cal N}=1$ susy.  For all the other cases, when $k>3$,
 no component is left invariant, and supersymmetry is broken.

 We now turn to the  gauge theory description.
  As is well known, the  ${\cal N}=4$ theory can be described in ${\cal N}=1$ language as follows:  it  has three chiral multiplets and one vector multiplet.
  A $U(1)$  subgroup of the full $SO(6)$ R-symmetry  is manifest in this description. Under it the scalar components of the three chiral multiplets, which correspond to the three coordinates, $Z^1,Z^2,Z^3$,  have R-charge $2/3$, their fermionic partners have R-charge $-1/3$ and the gaugino has charge $+1$. Besides the $U(1)_R$ symmetry this description also makes an additional $SU(3)$ subgroup of the $SO(6)$ R-symmetry group manifest.
  The three chiral superfields transform as a triplet of the $SU(3)$. The theory has a superpotential which is trilinear in the chiral superfields and which is also manifestly $SU(3)\times U(1)_R$ invariant.
  The $SU(3)$ symmetry we have identified in this way in fact corresponds to the $SU(3)$ isometries of the base $CP^2$ in the gravity description while the $U(1)_R$ corresponds to continuous shifts along the fiber.

  It then follows that the  rotation $R_{({2\pi\over k})}$ in eq.(\ref{defrot}) acts with a phase $e^{{3\pi i \over 2k}Q }$ on the fields of the ${\cal N}=4$ theory,
  where $Q$ is the R-charge of the field. Thus the full action of the generator $\alpha$ of eq.(\ref{genzn}) is by a phase
  $e^{{3\pi i \over 2k} Q} (-1)^F$. As an aside, note that the three fermionic partners of the scalars and the gaugino lie in a $ 4$ dimensional spinor representation of $SO(6)$. The transformation, eq.(\ref{transformsp}) follows from this.  Now to determine the resulting gauge theory,
  after the orbifold projection, we also need to embed the ${\mathbb Z}_{k}$ discrete symmetry in the $SU(N)$ gauge symmetry of the ${\cal N}=4$ theory. Here we take $N=nk$.
   And on the $N$ dimensional fundamental representation of $SU(N)$ we take the generator of the ${\mathbb Z}_{k}$ symmetry  to act in a block diagonal fashion as:
  \be
  \label{actfunds}
  ([1]_{n\times n}, [e^{2\pi i \over k}]_{n\times n}, [e^{4 \pi i \over k}]_{n\times n}\cdots, [e^{2(k-1) \pi i \over k}]_{n \times n}),
  \ee
 so that it multiplies each  $n\times n$ subspace by a $k$th root of unity. The action on any other representation  follows from this.  This is just the simplest example of the general procedure described in \cite{Douglas} for computing the
 spectrum of D-branes at orbifold singularities.

  Fields which survive in the orbifold theory are invariant under the simultaneous action by $e^{{3\pi i\over 2k} Q} (-1)^F$ and the  action on the gauge indices. It is easy to see that the orbifold projection breaks the $SU(N)$ gauge symmetry to $SU(n)^k$
resulting in a $k$ node quiver.\footnote{There are also extra $U(1)$ factors which we are not being careful about.}
The scalars and fermions give rise to bi-fundamental matter. In particular, for
$k>3$, there are no  fermions  which transform in the adjoint representation of $SU(n)^k$, and thus no gauginos. This shows that  supersymmetry is broken.

Let us give the resulting spectrum in full detail for the case $k=5$.
The $3$ complex scalars give rise to:
\begin{equation}
\label{scalarmattercontentn4}
\begin{array}{c|ccccc}
    & SU(n)_1 & SU(n)_2 & SU(n)_3 &  SU(n)_4 & SU(n)_5\\ \hline
 Q^i_1 & \Yfund  & \overline{\Yfund}  & 1       &  1 &1 \\
  Q^i_2 & 1   & \Yfund  & \overline{\Yfund}  &  1 & 1\\
 Q^i_3 & 1 & 1 & \Yfund  & \overline{\Yfund} & 1 \\
 Q^i_4 & 1 & 1 & 1 & \Yfund & \overline{\Yfund} \\
 Q^i_5 & \overline{\Yfund} & 1 &1 &  &  \Yfund
 \end{array} \nonumber
\end{equation}
Here the superscript $i$ takes three values,  $i=1,2,3,$.  The fields $Q_m^i, m=1 ,\cdots 5,$ arise from the complex scalar $Z^i$ in the ${\cal N}=4$ theory.
The fermions give rise to :
\begin{equation}
\label{fermionmattercontentn4}
\begin{array}{c|ccccc}
   & SU(n)_1 & SU(n)_2 & SU(n)_3 &  SU(n)_4 & SU(n)_5\\ \hline
\psi^i_1 & \Yfund  & 1 &  \overline{\Yfund}         &  1 &1 \\
  \psi^i_2 & 1      & 1 &   \Yfund  &  1 & \overline{\Yfund} \\
 \psi^i_3  & 1 & \overline{\Yfund} & 1 & 1 & \Yfund \\
 \psi^i_4 & 1  & \Yfund & 1 & \overline{\Yfund} & 1  \\
 \psi^i_5 & \overline{\Yfund} & 1  & 1 &\Yfund &  1 \\
 \lambda_1 & \overline{\Yfund} & \Yfund & 1 & 1 & 1 \\
 \lambda_2 & 1 & \overline{\Yfund} & \Yfund & 1 & 1 \\
 \lambda_3 & 1 & 1 & \overline{\Yfund} & \Yfund & 1 \\
 \lambda_4 & 1  & 1 & 1  & \overline{\Yfund} & \Yfund \\
 \lambda_5 & \Yfund & 1 & 1 & 1 & \overline{\Yfund}
 \end{array} \nonumber
\end{equation}
The $\psi^i_m, i=1, 2,3,$ fermions arise from the  fermions in the
 chiral multiplets (in ${\cal  N}=1$ language),  while the $\lambda_m$ fermions arise from the gaugino.

For those who find quiver diagrams more useful, the quiver summarizing this field content is given below.

\begin{figure}[h]
\begin{center}
\psfrag{A12}[cc][][1]{$A_{1,2}$}
\psfrag{B12}[cc][][1]{$B_{1,2}$}
\includegraphics[width=5cm]{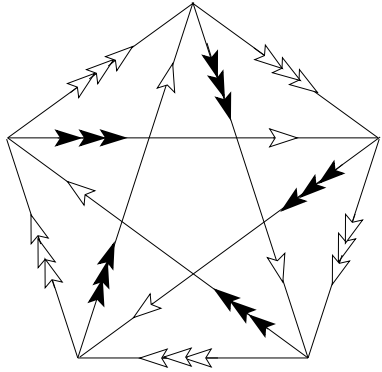}
\caption{Quiver diagram of the $k=5$ case.  White arrows denote fermions, and black arrows denote scalars.  We thank the authors of \cite{DKRone} for permission to reproduce this figure.}
\label{quiver_Z5}
\end{center}
\end{figure}

\subsection{Operator analysis}
Having understood the matter content of the field theory dual to  the orbifold   we can now investigate  whether there are relevant operators which would destabilise the throat. In the gravity picture these correspond to modes which would grow exponentially fast in the radial direction. In particular we are interested in relevant operators which are singlets under the $SU(3)\times U(1)_R$ global symmetry.

For this purpose it is useful to examine first how the  relevant operators in the  ${\cal N}=4$ theory transform under the $SU(3)\times U(1)_R$ symmetry.
Let us start with single trace gauge invariant operators.

The ${\cal N}=4$ theory has three kinds of operators which are bilinears in the scalar:

1) $Tr(Z^iZ^j)$ : These have dimension $2$. They transform like a ${\bf 6}$ of $SU(3)$ and carry charge $4/3$ under $U(1)_R$.  Thus they are not singlets under $SU(3)\times U(1)_R$. The operators,
$Tr(\bar{Z}_i \bar{Z}_j),$ which are  complex conjugates transform in the complex conjugate representation under the global symmetries and are also not singlets.

2) $Tr(Z^i\bar{Z}_j)-{1\over 3} \delta ^i_j Tr(Z^i\bar{Z}_i)$: These also have dimension $2$.  They are
 singlets under the $U(1)_R$ but transform like an $\bf{8}$ of $SU(3)$ and are therefore  not singlets under the global symmetry.

3)$Tr(Z^i \bar{Z}_i)$: This operator is a singlet.  However it has an anomalous dimension which goes like $\Delta \sim (g_s N)^{1/4}$ and thus is much bigger than unity in the large 't Hooft coupling limit. It is therefore not relevant.

In the orbifold theory there are also scalar bilinears which arise from the $Q^i_m$ fields and their complex conjugates. However these operators inherit their
 $SU(3) \times U(1)$ quantum numbers  and also their anomalous dimensions (to leading order in $N$) from the ${\cal N}=4$  theory. \footnote{This  is consistent with the fact that in the sugra approximation the mass of  invariant KK modes is left unchanged by the  orbifolding procedure.} Thus we conclude that there are no scalar bilinears in the orbifold theory which are global singlet relevant operators (GSROs).

The discussion above brings out one of the central points of the paper, so it is worth emphasising in more general terms.
At strong coupling (large 't Hooft coupling) in the supersymmetric parent theory, only protected operators have anomalous dimensions of order unity; these operators are charged under the global symmetries of the parent theory and thus are not GSROs. If we can arrange for a sufficiently big
subgroup of the global symmetry group to be  preserved by the daughter orbifold theory, it too will not contain any GSROs. In particular it was  vital in the example above that the operator $Tr(Z^i\bar{Z}_i)$ obtains an anomalous dimension bigger than $4$ at strong coupling in the ${\cal N}=4$ theory. This ensures that the daughter theory does not suffer from a hierarchy problem even though it has elementary scalars and no supersymmetry! In contrast, at weak 't Hooft coupling, the operator $Tr(Z^i\bar{Z}_i)$ has (approximately) dimension $2$ and is thus relevant.
It would be generated at the cut-off and destabilise the orbifold field theory. The importance of this large anomalous dimension at
strong coupling was emphasised in \cite{Strassler}.

Continuing with our discussion of possible GSROs,  one class of  dimension three operators in the ${\cal N}=4$ theory arise from Lorentz invariant, fermion bilinears.  Denoting the three matter fermions by $\psi^i, i=1, \cdots 3$ and the gaugino by $\lambda$,  there are three operators of this type:

1)$Tr(\psi^i\psi^j)$: R-charge $-{2\over 3}.$

2)$Tr(\lambda \lambda)$: R charge $2.$

3)$Tr(\psi^i\lambda)$: R charge $2/3.$

Thus none of these are global singlets. As a result no fermion bilinear global singlets arise in the daughter theory. Actually there is an even quicker argument which one can use in this case: the daughter theory has fermions in only bi-fundamental representations (no adjoints). It is easy to see that no Lorentz invariant, gauge invariant fermion bilinears can be made from these.

Additional dimension 3 operators in the ${\cal N}=4$ theory arise from scalar trilinears.
Each scalar in the trilinear can be one of the $Z^i$ or the $\bar{Z}_i$ fields. However since the $Z^i$s have R-charge $2/3$ it is easy to see that no such trilinear combination can be $R$-charge neutral. Thus, no GSROs can arise from these operators either.

Having discussed all possible single trace GSROs, let us now turn to double trace operators.  The smallest dimension of a single trace gauge invariant in the ${\cal N}=4$ theory is  $2$. Since to   leading order in $N$ the anomalous dimensions of double trace operators are simply the sum of their   single trace constituents,  it follows that any double trace operator must have at least dimension $4$ and can therefore at most be marginal. This completes our discussion of possible GSROs in the orbifold theory. We see that there are no such operators and thus the throat in the dual gravity description is stable.

Note that we have not discussed possible global singlet marginal operators (GSMOs). These are operators whose dimension is $4$, up to small corrections.
 Such operators are in fact present in the parent theory and thus also arise in the daughter theory. One example is a double trace operator made out of single trace scalar bi-linears. No symmetry prevents these operators and they will be generated by radiative effects even if one sets them to zero in the ultraviolet. However since these operators  are marginal their presence will not destabilise the hierarchy, which is our main concern here. In fact such operators can play an important role in ensuring the vacuum stability of the orbifold theory.
To leading order in $N$ the orbifold theory has flat directions - these are inherited from the ${\cal N}=4$ theory and correspond to ${\mathbb Z}_{k}$ symmetric displacements of the $D3$-branes along the Coulomb branch.  These flat directions will be lifted by quantum effects and could potentially lead to Coleman-Weinberg type run-away instabilities.
Since the GSMOs will be radiatively generated anyways, one might as well  add them to the tree level Lagrangian with appropriately small coefficients.
These coefficients (with sign) can be chosen to lift at least some of the flat directions. We will  not pursue a complete analysis of the resulting stability of these orbifold theories here,  see \cite{Strassler}
 for some discussion.
In the subsequent section an example is constructed in detail based on
an orbifold of the Klebanov Strassler theory. In this example we will see that there is no vacuum instability.

Let us end with one final comment. We have used the continuous symmetry
$SU(3)\times U(1)_R$ to prevent relevant operators. However it is well known that
realistic compacifications of string theory, like Calabi-Yau compactifications,  do not give rise to isometries.  So one might be worried that after compactification these isometries will be broken and the relevant operators cannot be prevented. However, one can easily  construct examples of Calabi-Yau manifolds with  unbroken discrete symmetries. A moderately big discrete symmetry can often suffice to prevent
operators of dimension $<4$. This will be illustrated in detail in the example based on the Klebanov-Strassler theory below.

\section{An Overview of KW and KS}

While the discussion of ${\cal N}=4$ orbifolds already provides a wide class of non-SUSY theories without GSROs,
it is useful to study a single example in more detail.  One would explicitly like to construct a compactification preserving enough
symmetries to protect the hierarchy, and also provide a more detailed picture of the emergent IR physics.
For these purposes, we find it useful to study an example based on D-branes at the conifold. Below we construct such an example based on an orbifold of the Klebanov-Strassler theory. 
In this section we  review some essential features of the Klebanov-Witten  and the Klebanov-Strassler   theories, \cite{KW}, \cite{KS},  and then turn to the non-supersymmetric orbifold in the section which follows.

\subsection{The Klebanov-Witten (KW) Theory}

The Klebanov-Witten theory is obtained by placing $D3$ branes at the tip of a conifold. The resulting  gauge group is $SU(N)\times SU(N)$,
with chiral multiplet  matter fields  $A_i, B^j, i=1,2; j=1,2$ transforming as
$(\Yfund, \overline{\Yfund})$ and $(\overline{\Yfund}, \Yfund )$ respectively under the gauge symmetries. The non-anomalous global symmetries include a Baryonic symmetry,  $U(1)_B$, under which $A_i$ have charge $+1$ and
$B^j$ have charge $-1$, and a $U(1)_R$ R-symmetry, under which the fields $A_i,B^j$ have charge $1/2$.
The theory also has an $SU(2)\times SU(2)$ flavor global symmetry. The fields
$A_i, i=1,2$ transform as a doublet of the first $SU(2)$ and the $B^j ,j=1,2$ as a doublet of the second $SU(2)$.
There is in addition a discrete  ${\mathbb Z}_2$
symmetry which we will refer to as ${\mathbb Z}_{2 {\rm exchange}}$ below. This  involves the exchange  $A_i\leftrightarrow B^j$ accompanied by complex conjugation.

 The dual gravity description of this field theory involves type $IIB$ string theory on $AdS_5\times T^{1,1}$. The $U(1)_R$ symmetry corresponds to an isometry of the $T^{1,1}$ manifold. $T^{1,1}$ has an $O(4)$ symmetry group. The $SU(2)\times SU(2) $ flavor symmetry is a subgroup of $O(4)$.
The ${\mathbb Z}_{2 {\rm exchange}}$  mentioned above corresponds to acting by a reflection  element  of $ O(4)$ (with determinant $-1$),
accompanied by $\Omega(-1)^{F_L}$ where $\Omega$ stands for orientation reversal on the world sheet.
Under  ${\mathbb Z}_{2 {\rm exchange}}$, the two forms $B_2,C_2,$ have odd intrinsic parity.


Of special importance to us, as was discussed in the previous section, are relevant operators, in particular operators with dimension less than $4$, which are Lorentz scalars and singlets under the global symmetries. We turn to studying these next.
The spectrum of KK modes for the KW theory was calculated by Ceresole  and collaborators \cite{Ceresole}.
 It is easy to read off the spectrum of all  relevant operators from their work.
 We will skip some of the details here, and only discuss operators
 which are singlets under the $SU(2)\times SU(2)$ global symmetry.\footnote{
 We thank A. Dymarsky and O. Aharony for  discussions in this regards, and A. Dymarsky for informing us of some
 minor corrections to the results in \cite{Ceresole}.}    Operators which are ${\it not}$ $SU(2) \times SU(2)$ singlets, can
 be naturally forbidden by preserving a large enough subgroup of $SU(2) \times SU(2)$ in the non-supersymmetric
 model we construct in \S4.

 There is only one  operator of dimension $2$ which is a singlet under the $SU(2)\times SU(2)$ global symmetry.
 It is $Tr(|A|^2-|B|^2)$, which is the scalar component of the
 $U(1)_B$ current multiplet.   Here we are being a bit schematic  -- the trace is over the colour degrees of freedom, and also the $SU(2)\times SU(2)$ flavour indices.
 Note that this operator is ${\it odd}$ under the ${\mathbb Z}_{2 {\rm exchange}}$ symmetry, and is therefore not a
 singlet of the full global symmetry group.  This means it will be important for us to maintain the
 ${\mathbb Z}_{2 {\rm exchange}}$ symmetry in our non-supersymmetric construction.

At dimension $3$, there are three operators which are singlets under the
$SU(2)\times SU(2)$ global symmetry. Two of these are the gaugino bilinears of the two gauge groups, $Tr(\lambda \lambda)$.
However these carry $R$-charge $2$ under the $U(1)_R$.
The third operator is   $Tr(A_1B_1A_2B_2 - A_1B_2A_2B_1)$.
 This is the same operator which appears in the superpotential, but here the operator we
 are considering only contains   the scalar components of the chiral superfields  $A,B$.
 Once again this operator has R-charge $2$.
 Thus there are no single trace global singlets of dimension $<4$  in this theory.

 Next we turn to double trace operators. It is easy to see that there is in fact
 one double trace operator which is a global singlet  in the theory with dimension $3$. It is
 given by,
  $Tr(AB)\overline{{Tr(AB)}}$.
 Here, the bar in the second term indicates the complex conjugate of $Tr(AB)$. The
 trace is over colour indices and the $SU(2)\times SU(2)$ flavour indices have been contracted between the two single trace operators to result in a
 singlet under the full flavour group. In the large $N$ limit the dimension
 of a double trace operator is given by the sum of the dimensions of the
 two single traces. Since $Tr(AB)$ has dimension $3/2$ it then follows that this double trace operator has dimension $3$.
 We will describe how this operator is eliminated in our non-supersymmetric construction in \S4, once we have
 provided the relevant quiver diagram.
 The other dimension 3  $SU(2)^2$ singlet double-trace operators, of
 the form $Tr(AB) Tr(AB)$ (and their conjugates),
 have R-charge 2, and can be forbidden by maintaining a large enough subgroup of $U(1)_R$.

 In conclusion, in the KW theory there is only one operator with dimension $<4$
 which is a global singlet.  It is the double trace operator $Tr(AB) \overline{{Tr(AB)}}$.

 Before proceeding further it is worth emphasizing a few  important points:

\medskip
\noindent
$\bullet$
 First, the field theory above has scalars in it.
 At ${\it weak}$ coupling this means that there is ${\it always}$ a global singlet of dimension $2$;
 in the KW theory it is of the form $Tr(|A|^2+|B|^2)$. However, we see that at strong 't Hooft coupling this operator
 acquires a large anomalous dimension, and in fact does not correspond to a SUGRA mode, but rather to a string mode. This is akin to what we saw in the ${\cal N}=4$ theory above,  where the scalar bilinear which is an $SO(6)$ singlet acquires a big anomalous dimension at strong coupling. In the KW case too this feature  plays an important role in ensuring the absence of GSRO's.

 \medskip
 \noindent
 $\bullet$
 Second, one may worry that $U(1)$ symmetries which
 are not $R$ symmetries are always problematic with regards to a stable hierarchy.
 From the representation theory of ${\cal N}=1$ superconformal symmetry it is known that the  multiplet containing a $U(1)$ current must also contain a scalar of dimension $2$. This scalar must be a singlet under all the continuous global symmetries and thus is in general problematic.\footnote{We thank K. Agashe and R. Sundrum for  stressing this concern.} We see from the discussion above that this conclusion can be sometimes avoided. The KW theory has a $U(1)_B$ current as was mentioned above, and in fact the  dimension $2$ scalar we found above,
 $Tr(|A|^2-|B|^2)$, is the partner of the   $U(1)_B$ current. However we see that
 the theory posses in addition a ${\mathbb Z}_{2 {\rm exchange}}$ discrete symmetry which does not commute with Baryon number. This symmetry prevents the scalar partner of the Baryonic current from destroying the hierarchy.  A similar argument could work more generally for a Baryon current in a theory which has charge conjugation symmetry.

\medskip
\noindent
$\bullet$
 Finally, we have been considering a non-compact situation above, where the
 AdS throat extends to infinity in the UV and there is no dynamical $4$ dimensional gravity. For added realism
 we should consider embedding the KW throat in
 a compact Calabi-Yau manifold. Now it is well known that there are no continuous isometries in compact
 Calabi-Yau manifolds with sufficiently generic holonomy. Thus one would expect that once the
 KW throat is glued into the compact Calabi-Yau space, relevant operators which are not singlets
 under the global symmetries will also be induced in the theory in the ultra-violet. Such operators will then destroy the hierarchy.
 To avoid this conclusion, we can consider situations where a sufficiently large discrete subgroup of the $SU(2)\times SU(2)\times U(1)_R  \times {\mathbb Z}_{2 {\rm exchange}}$ symmetry\footnote{This notation is a bit loose. The ${\mathbb Z}_{2 {\rm exchange}}$
  does not commute with the two $SU(2)$'s.  In fact it exchanges them, and also reverses the $U(1)_B$ charges. So the full symmetry group is a semi-direct product, rather than a product.} is preserved by the Calabi Yau manifold. This could then
 suffice to prevent  relevant operators
 (with dimension $< 4$) from being induced, even after coupling the (approximate) CFT to quantum gravity. We provide an explicit example in \S4 showing that this can indeed happen.

\subsection{The Klebanov-Strassler (KS) Theory}

 The KS theory is a  deformation of  the KW theory obtained by taking the two
 gauge groups to have unequal rank (see the Figure below). The resulting
 gauge theory  has gauge group $SU(N+M)\times SU(N)$ with matter fields $A_i,B^j$.  The $U(1)_B$ Baryonic symmetry mentioned in the discussion of the KW theory continues to be non-anomalous in this theory.
  The $U(1)_R$ symmetry is now anomalous but a ${\mathbb Z}_{2M}$  subgroup survives as a non-anomalous discrete symmetry of the Lagrangian. The theory
 has an $SU(2)\times SU(2)$ global symmetry which acts on the $A_i,B^j$ fields as in the KW case. Also the ${\mathbb Z}_{2 {\rm exchange}}$ discrete symmetry  continues to be a symmetry in  the KS case.
The theory undergoes a duality cascade under RG flow. At each step in the cascade,
the rank $N$ changes by
 $N\rightarrow N-M$. In the deep infrared the ${\mathbb Z}_{2M}$ R symmetry is broken spontaneously to a ${\mathbb Z}_2$ subgroup. In the far ultraviolet the rank
 $N \rightarrow \infty$, and the theory approaches the KW case.

\begin{figure}[h]
\begin{center}
\psfrag{A12}[cc][][1]{$A_{1,2}$}
\psfrag{B12}[cc][][1]{$B_{1,2}$}
\includegraphics[width=5cm]{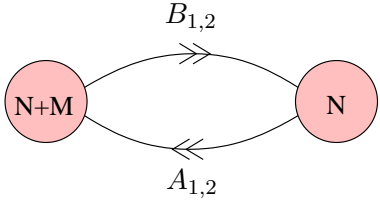}
\caption{Quiver diagram for the KS theory.}
\label{quiver_conifold}
\end{center}
\end{figure}

In the gravity description, the  parameter $M$ corresponds to Ramond-Ramond three-form flux $F_3$ which is turned on along a non-trivial $3$-cycle.   The $3$-form flux results in the $U(1)_R$ symmetry being broken to ${\mathbb Z}_{2M}$ \cite{Ouyang}. Since this flux is invariant under the ${\mathbb Z}_{2 {\rm exchange}}$ discrete symmetry which acts as a reflection combined with $\Omega (-1)^{F_L}$,
the symmetry remains unbroken. The back-reaction due to the additional three-form fluxes (SUSY requires that NS-NS flux
$H_3$ is also turned on) grows in strength in the infrared and results in a deformation of the conifold. This deformation of the conifold breaks the ${\mathbb Z}_{2M}$ symmetry to a  ${\mathbb Z}_2$ subgroup in the infrared. In the far ultra-violet the
effects of the three-form flux are negligible compared to that of the $5$-form
and the geometry approaches that of the $AdS_5\times T^{1,1}$ case,  with logarithmically small corrections.

Next, we turn to a discussion of operators with dimension $<4$ in the KS theory.
Since, as was mentioned above,  this theory approaches the KW theory in the ultra-violet up to logarithmic corrections, the dimension of operators in the UV in the KS case can be obtained directly from our earlier discussion of the KW case, up
to small corrections.
There is one important difference:
 while the $SU(2)\times SU(2)$ and the $ {\mathbb Z}_{2 {\rm exchange}}$
symmetry are preserved in the KS case, the $U(1)_R$ symmetry is broken to a
${\mathbb Z}_{2M}$ subgroup in the KS theory.\footnote{It is further broken spontaneously from ${\mathbb Z}_{2M}$ to
${\mathbb Z}_2$ in the deep infrared.  The spontaneous breaking
of ${\mathbb Z}_{2M} \to {\mathbb Z}_2$ is not a worry in the context
of perturbations that can destroy the throat.
Operators protected by the ${\mathbb Z}_{2M}$ symmetry might be induced in the IR once the symmetry breaks to ${\mathbb Z}_2$, but this will not destabilize the hierarchy.}
Thus the global symmetry group available to us in the KS case is smaller. Operators which are of dimension $<4$ and which are not singlets under this smaller global symmetry group can destabilise the hirarchy.
Note that the breaking of $U(1)_R$ symmetry to ${\mathbb Z}_{2M}$ occurs due to an anomaly,
and is supressed in $M/N$. However, in  realistic compactifications, one does not expect (due to tadpole cancellation conditions etc.)  that   exponentially large values of $N$ are allowed (even if they were aesthetically tenable);
thus this supression by itself is not enough to ensure the stability of an exponentially large  hierarchy.

Looking through the list of operators of dimension $<4$ discussed above in the KW theory again, we see that among the single trace operators the $U(1)_R$ symmetry
was important in protecting the hierarchy from the three dimension $3$ operators, all of which have
 $R$-charge $2$. While the $U(1)_R$ symmetry is broken to a ${\mathbb Z}_{2M}$ subgroup in the KS case, for $M>1$ this is still a big enough residual symmetry to prevent these operators from being induced.   Among the double trace operators the $U(1)_R$ symmetry was important for operators of the form, $Tr(AB) Tr(AB)$.
 These have $R$ charge $2$ also and therefore they will also be forbidden by the surviving ${\mathbb Z}_{2M}$ symmetry. This only leaves the double trace non-holomorphic
  operator of the form, $Tr(AB) \overline{{Tr(AB)}}$, which is a singlet under all the global symmetries. It is a GSRO in the KS theory  as well.  Thus, we see that in the KS case, as in  the KW theory,
 the only
 global singlet
 operator with dimension $<4$ is the double trace operator $Tr(AB) \overline{{Tr(AB)}}$.

   \section{The Non-SUSY Orbifold}
   We are now ready to consider the breaking of SUSY. This will be accomplished
   by constructing an orbifold. Our real interest is in the KS theory, but as in the discussion above it will also be useful to discuss the KW theory as we proceed.

   The orbifold group must involve the unbroken symmetries of the KS solution.
Since we want the resulting orbifold to break SUSY, it must involve the $R$ symmetry group.
We have seen above that the KS theory only preserves an unbroken  ${\mathbb Z}_2$ subgroup of the underlying non-anomalous ${\mathbb Z}_{2M}$ R-symmetry group.\footnote{Note that on the $A_i$, $B^j$ fields,  which
carry R-charge $1/2$, the generator of this subgroup acts with a phase of $i$.  This is consistent
with the symmetries of the deformed conifold, since the $z_i$ variables appearing in (\ref{con}) are
bilinears in $A, B$, and taking $z_i \to - z_i$ is a symmetry even after deforming the conifold.}
The simplest possibility, then, which leaves the $SU(2)^2$ symmetry untouched, is to consider the orbifold group to be this ${\mathbb Z}_2$ subgroup, possibly
combined with discrete subgroups of the $U(1)_B$ symmetry.

We will choose to accompany the ${\mathbb Z}_2$ R-transformation above with an action of $U(1)_B$ which
rotates the $A$ fields by $i$ and the $B$ fields by $-i$.
The result is that we quotient by a  ${\mathbb Z}_2$ R-symmetry under which the scalar components of the chiral fields transform as:
\be
A\rightarrow -A, B\rightarrow B~.
\ee
This means the fermionic partners transform by:
 \be
 \psi_A\rightarrow \psi_A, \psi_B\rightarrow -\psi_B~.
 \ee
  The gauginos of the two gauge groups transform, as usual  under a ${\mathbb Z}_2$ R symmetry, as
  \be
  \lambda \rightarrow -\lambda~.
\ee

Before proceeding it is important to clarify one point.
A different ${\mathbb Z}_2$ symmetry can be defined by combining the ${\mathbb Z}_2$ action discussed above
with $(-1)^F$, where $F$ is spacetime fermion number.
This new ${\mathbb Z}_2$ symmetry has the same action in spacetime but it acts oppositely on spacetime fermions.  As a result it turns out to preserve supersymmetry. Under it, $(A,\psi_A)\rightarrow -(A,\psi_A), (B,\psi_B)\rightarrow (B,\psi_B)$,
while the gauginos are invariant, $\lambda \rightarrow \lambda$.
Orbifolding by this ${\mathbb Z}_2$ gives rise to a SUSY-preserving quiver theory, which has been discussed in \cite{uranga,morrison}. This quiver is different from the one which we will obtain after orbifolding by the SUSY-breaking ${\mathbb Z}_2$ symmetry described in the previous paragraph. This will become clear when we discuss the matter content of the SUSY-breaking case in some detail below.

Now let us continue with our  discussion of  the SUSY-breaking ${\mathbb Z}_2$ orbifold in more detail.
 Note that the ${\mathbb Z}_2$ action has no fixed points in the dual IIB gravity description.
The unwarped conifold is described by the locus
\be
\label{con}
z_1z_2 - z_3z_4=0
\ee
where each of the $z_i$ coordinates can be expressed as  a bilinear product of one of the $A$ and
one of the $B$ fields. From the transformations of the $A$ and $B$ fields given above, it follows that  under the
${\mathbb Z}_2$  -symmetry
\be
\label{trnz}
z_i\rightarrow -z_i~.
\ee
This appears to have a fixed point at the (singular) tip of the conifold $z_i = 0$.  However,
as in \S3\ of  \cite{orbCFT}, this is not the case in the near-horizon limit of N D3-branes probing this
geometry.  We can think of  (\ref{con}) as a cone with $S^3 \times S^2$ base over a radial
direction $r$, where the $S^3 \times S^2$ shrinks at $r=0$.  The near-horizon limit chooses a slice of definite size for
the $S^3 \times S^2,$ yielding the geometry $AdS_5 \times T^{1,1}$.  The $T^{1,1}$ is
a slice of the cone at some definite $r > 0$ in the above description, and the fixed point
at $z_i=0$ does not survive the near-horizon limit.

Because the ${\mathbb Z}_2$ acts freely on $AdS_5 \times T^{1,1}$, it is easy to determine the states in the theory  after orbifolding. At the SUGRA level these  correspond
to KK modes which are invariant under the orbifold symmetry.  In particular, there are no additional twisted sector states we need to worry about.  (There are of course twisted sector string states,
but these correspond to operators of sufficiently high dimension that they are of no concern to us).

On the gauge theory side, in determining the quiver gauge theory which arises after orbifolding, it is again useful to first consider the KW case.
To determine the quiver theory   we must also embed the action of the ${\mathbb Z}_2$ in the $SU(N)\times SU(N)$ gauge group.
 Here we consider the standard embedding (for a discussion of such orbifolds of D-brane theories
 in general terms, see \cite{Douglas}). Take $N=2n$. In the fundamental representation of $SU(N)$ --
 in terms of $N\times N$ matrices with unit determinant -- the ${\mathbb Z}_2$ symmetry is given by $(I_{n\times n},-I_{n\times n})$. This is the ${\mathbb Z}_2$ action in both of the $SU(N)$ subgroups.
 It is then easy to see that the resulting quiver has $4$ nodes, each corresponding to a $SU(n)$ gauge group.
 The scalar fields transform as follows:
\begin{equation}
\label{smattercontent}
\begin{array}{c|cccc}
    & SU(n)_1 & SU(n)_2 & SU(n)_3 &  SU(n)_4 \\ \hline
 Q_1 & \Yfund  & \overline{\Yfund}  & 1       &  1 \\
  Q_2 & 1   & \Yfund  & \overline{\Yfund}  &  1 \\
 Q_3 & 1 & 1 & \Yfund  & \overline{\Yfund}  \\
 Q_4 & \overline{\Yfund} & 1 &1 &  \Yfund
 \end{array} \nonumber
\end{equation}
Here $Q_1,Q_3$ arise from the field $A_1$ and $Q_2,Q_4$ from the field $B_1$.
Similarly there are scalars which arise from $A_2,B_2$ as well, giving rise to two copies of
this scalar spectrum.

 The fermionic fields transform as follows:
 \begin{equation}
\label{fmattercontent}
\begin{array}{c|cccc}
    & SU(n)_1 & SU(n)_2 & SU(n)_3 &  SU(n)_4 \\ \hline
 \psi_1 & \overline{\Yfund}  & \Yfund  & 1       &  1 \\
  \psi_2 & 1   & \overline{\Yfund}  & \Yfund  &  1 \\
 \psi_3 & 1 & 1 & \overline{\Yfund}  & \Yfund  \\
 \psi_4 & \Yfund & 1 &1 &  \overline{\Yfund}
 \end{array} \nonumber
\end{equation}
Here $\psi_1,\psi_3$ descend  from the fermionic partner of $B_1$, and $\psi_2,\psi_4$
descend from the fermionic partner of $A_1$ in the parent theory. Similarly there are fermions that descend  from the fermionic partners of $B_2,A_2$.  So again, we get two copies of this
fermionic spectrum.

 Additional matter also arises from the gauginos in the parent theory.
 They give rise to bi-fundamental fermionic matter which transforms as follows:
 \begin{equation}
\label{smattercontent}
\begin{array}{c|cccc}
    & SU(n)_1 & SU(n)_2 & SU(n)_3 &  SU(n)_4 \\ \hline
 \lambda_1 & \Yfund  & 1& \overline{\Yfund}         &  1 \\
  \lambda_2 &  \overline{\Yfund}  & 1 & \Yfund  &  1 \\
 \lambda_3 & 1 & \Yfund  & 1 & \overline{\Yfund}  \\
 \lambda_4 & 1 & \overline{\Yfund} & 1  &  \Yfund
 \end{array} \nonumber
\end{equation}

It is clear from the matter content above that the resulting quiver theory breaks supersymmetry. For example there are no fermions in the adjoint representation of the
quiver gauge group, and thus no possible gauginos in this theory.

 The discussion in the KS case is essentially similar. We start with the gauge group $SU(2n+2m)\times SU(2n)$ and consider the standard embedding of the ${\mathbb Z}_2$
 action in the two gauge groups. This results in a four node quiver with gauge group $SU(n+m)\times SU(n)\times SU(n+m)\times SU(n)$ and matter content consisting of bifundamentals which form a quiver diagram identical to the KW case.

 Let us now turn to the global symmetries.
 The $SU(2)\times SU(2)$ global symmetry is still preserved in the non-supersymmetric
 quiver theory:
 the descendants of the $A$ fields transform as a doublet under the first $SU(2)$ and those of the $B$ fields as a doublet of the second $SU(2)$. The $U(1)_B$ is also preserved, with the descendants of the $A,B$ fields having charge $\pm 1$ respectively.  A ${\mathbb Z}_2$ symmetry analogous to the ${\mathbb Z}_{2 {\rm exchange}}$ symmetry in the KW/KS theories can be defined in the quiver theory. It can be described as follows.
 Exchange the descendants of the $A,B$ fields, along with exchanging the two
 $SU(n+m)$ groups with each other followed by charge conjugation. It is easy to see that this keeps the quiver diagram invariant and thus is a symmetry of the theory.
   Since it involves the exchange of the $A,B$ fields, this  ${\mathbb Z}_2$ symmetry also anti-commutes with Baryon number. We will refer to this symmetry are ${\mathbb Z}_{2 {\rm exchange}}$ below. The KS theory has a ${\mathbb Z}_{2M}$ R symmetry which is then spontaneously broken to a ${\mathbb Z}_2$ subgroup (so it is a symmetry in the UV of the solution, but is broken at the tip). The orbifold is obtained by identifying points in the KS geometry related to each other by this ${\mathbb Z}_2$
   action. In the orbifold  space a ${\mathbb Z}_{M}$ subgroup of the ${\mathbb Z}_{2M}$ R-symmetry still acts non-trivially and is a  global symmetry in the UV.
   Thus, the global symmetries of the orbifold theory are $SU(2)\times SU(2)\times U(1)_B
   \times {\mathbb Z}_{M} \times {\mathbb Z}_{2 {\rm exchange}}$.

 Now, we are ready to discuss the relevant operators in the orbifold theory. At the single trace level these operators will arise from the
 relevant operators of the parent theory. The dimension $2$ operators which arise
 from $Tr(|A|^2-|B|^2)$ in the parent theory are all odd under the ${\mathbb Z}_2$ symmetry defined above, which exchanges the descendants of the $A,B$ fields, and therefore are not singlets under the global symmetry group.
 The dimension three operators which arise all carry charge under the ${\mathbb Z}_{M}$  symmetry group (which survives as a symmetry from the underlying ${\mathbb Z}_{2M}$ R symmetry group) and, again, are not
 global singlets.
 This only leaves the possibility of double trace operators. However, it is easy to see that there are no global singlet double trace operators with dimension $<4$ that survive in the quiver theory, either.

 The point is that in the parent theory  the single trace operator $Tr(AB)$,  out of which the problematic
 double trace operator is composed,  is not invariant under the ${\mathbb Z}_2$ orbifold symmetry (since that symmetry takes $A\rightarrow  -A, B\rightarrow B$).  As a result there is no gauge invariant operator which arises in the daughter theory which is bilinear with one descendant from the $A$ and $B$ fields respectively.
 Without such a single trace operator no double trace operator can then arise.
  To get a gauge invariant single trace operator made out of the scalars  in the daughter theory one needs two descendants from the $A$ and two from the $B$ fields respectively. Such an operator can be thought of as arising from an operator of the type $Tr(ABAB)$ in the parent theory. It has dimension $3$, thus a double trace operator made out of two such single trace operators would have dimension $6$, and would be irrelevant.

  The conclusion is that the orbifold theory has no operators, either single trace or double trace, which are singlets under the global symmetries and which have dimension $<4$.

This ensures that our first aim is met: we have constructed non-compact non-supersymmetric warped backgrounds with
a stable throat geometry (or equivalently, a stable hierarchy of scales).

 One might worry that coupling this field theory to quantum gravity would be problematic.
 After all, quantum gravity famously abhors at least continuous global symmetries.  We display,
 in the next section, a compact embedding of this throat geometry, where the compactification
 preserves a sufficiently large subgroup of the global symmetry group to protect the hierarchy.

Let us end with a comment about vacuum stability.
In the  KS theory, before orbifolding, if we take $N=kM$,  the  duality cascade ends in the IR with a   ${\cal N}=1$ $SU(M)$ gauge theory, with no additional matter.
This theory confines and has a mass gap. The vacuum of this theory is stable. In particular there are no massless scalars, which could be rendered unstable due to quantum effects.  In the orbifold theory, with $M=2m$ the end point is a two node quiver with gauge group $SU(m)\times SU(m)$, and a pair of bifundamental fermion fields which transform like $(\Yfund, \overline{\Yfund})$ and $(\overline{\Yfund}, \Yfund)
$ respectively. This theory inherits a mass gap from the parent theory,and thus does not suffer from a vacuum instability. In fact this agrees with what one would expect from the gauge theory analysis. Any one of the two $SU(m)$ gauge theories
has   $m$ flavours in the fundamental representation. It should confine and in fact exhibit chiral symmetry breaking, resulting in  a stable vacuum and a  mass gap.



 \section{Coupling to 4D Gravity}

Any model of the real world must incorporate 4d gravity, and this implies at least one interesting constraint on the previous discussion, which is the absence of continuous global symmetries
(see e.g. \cite{Banks}, \cite{Susskind} for general discussions). Therefore, we will relax the assumptions of the previous sections and assume that, in proving the absence of global singlet relevant
operators, we only have discrete symmetries at our disposal.  We will carry out the analysis for the theory of \S4, but presumably
one could easily find analogous constructions coupling the orbifolds of ${\cal N}=4$ to 4d gravity while preserving discrete symmetry
groups that forbid all relevant perturbations.

Inclusion of 4d gravity is achieved by cutting off the throat at some radius and gluing it into a compactification, thus making the 4d graviton a dynamical mode. The statement that there are no continuous global symmetries then corresponds to one of two possibilities: i) the compactification preserves the isometries of the throat, in which case the global symmetries are effectively gauged;  ii) the compactification breaks the isometries of the throat down to a (possibly trivial) discrete subgroup. When considering Calabi-Yau compactifications, as we will
do shortly, the second possibility is guaranteed:  Calabi-Yau manifolds (with sufficiently generic holonomy)
have no continuous isometries.   Our non-supersymmetric theory will be coupled to 4d gravity
 by taking an appropriate ${\mathbb Z}_2$ orbifold of a Calabi-Yau compactification with a conifold
 throat.  Therefore, we are guaranteed that our global group will be broken to at most discrete
 factors.  We must prove that there exist compact embeddings that preserve a sufficiently
 large discrete group, to forbid generation of all of the dangerous relevant operators.

In \S5.1 we find an explicit F-theory compactification which realizes a discrete subgroup of $G  = SU(2)\times SU(2) \times {Z}_{2M} \times {\mathbb Z}_{2 {\rm exchange}} $, and in \S5.2 we show this
global group is large enough to accomplish our task.

One could of course worry about the further steps involved in coupling to 4d quantum gravity: one must
show that the Calabi-Yau compactification can be stabilized appropriately, give rise to realistic
cosmology, etc.  These further steps are necessary in any attempt to embed any idea about
particle phenomenology in string theory, and are not special to our goal here of exhibiting composite Higgs models.   We will not pursue them in this note.

\subsection{An F-theory compactification}

Here we exhibit an F-theory compactification on an elliptically fibered four-fold which preserves a healthily large discrete subgroup of $G$. Consider the Weierstrass form
\be
y^2 = x^3 + f(z_i)~ x z^4 + g(z_i)~ z^6
\label{fiber}
\ee
where the $z_i$ are coordinates on the base $B$ of the elliptic fibration, not to be confused with  the $z$ appearing above. We take as our base $B$\footnote{We are inspired by seeking the simplest possible modification of the compact embedding of the conifold discussed in \S4\ of \cite{GKP}. That example doesn't suffice for our purposes.  It only preserves a
${\mathbb Z}_{2} \subset U(1)_R$, which
for instance isn't restrictive enough to forbid the dimension 3 gaugino mass operator $Tr~ \lambda \lambda$ from destroying the throat.} a complete
intersection of two quadrics in $P^5$.  If we label the defining equations of the base
$I_{1,2} = 0$, then the locus in moduli space we choose to work with is:
 \be
\label{base}
I_1 = \sum_{i=1}^{4} z_i^2 - t^2 z_5^2 + \epsilon^2 z_6^2,~~I_2 = z_1 z_2 + z_1 z_3 + z_1 z_4
+ z_2 z_3 + z_2 z_4 + z_3 z_4 + z_5 z_6~.
 \ee
This complete intersection is non-singular for generic values of $t$, $\epsilon$, which we
take to be real numbers.  As $t^2 \to 0$
or $\epsilon^2 \to 0$, there is a point of non-transversality (located respectively at $z_5=1$ with the other homogeneous coordinates vanishing, or $z_6= i$ with the other homogeneous coordinates vanishing).

Expanding in local coordinates around these singular points, we
see that the singularities are conifolds.  The collapsing $S^3$s can be seen to lie on the fixed point
loci of the involutions $z \to \bar z$ in the first case, and $z_{1,\cdots,4} \to \bar z_{1,\cdots,4}$ with
$z_{5,6} \to - \bar z_{5,6}$ in the second case.\footnote{In the Calabi-Yau orientifold we construct based on this example, one can see that the spheres are special Lagrangian; this is guaranteed for
fixed-point loci of antiholomorphic involutions with suitable properties \cite{math}.}
In particular, for non-zero but small
$t^2$, then, there is a deformed conifold singularity with a small $S^3$.  We shall use this
conifold throat associated with the singularity at $t^2 \to 0$ to build our approximately conformal
field theory.

To make the manifold (\ref{fiber}) Calabi-Yau, we should take $y \in 3L$, $x \in 2L$,
$f \in H^{0}(4L)$ and $g \in H^{0}(6L)$ where $L$ is the line bundle given by
$L = -K_{B}$ in terms of the canonical bundle of $B$.  In practice, for this model, we can think
of $f$ and $g$ as being polynomials of degree $8$ and $12$ in the coordinates of the
$P^5$.

Sen has given a general prescription for going to an orientifold locus of any fourfold
compactification \cite{Sen}.  Following his prescription, we wish to choose polynomials of
the special form

\be
f = C \eta(z_i) - 3h(z_i)^2, \quad g =  h(z_i) [ C \eta(z_i) - 2h(z_i)^2 ]~.
\ee
where $\eta$ and $h$ are of degree 8 and 4 respectively and $C$ is a constant.
For small (non-zero) values of $C$, the average coupling in the IIB string theory is weak.
The IIB theory lives on the base, eq.(\ref{base}).

A big discrete subgroup of $G$
is preserved by various simple choices of the data $\eta, h$.  For instance, we can take
 \be
\label{defeta}
\eta(z_i) = \sum_{i=1}^{4} z_i^{8} +  a z_5^{8} + z_6^8
\ee

\be
\label{defh}
h(z_i) = \sum_{i=1}^{4} z_i^4 + b z_5^4 +  z_6^4
\ee
for some tunable constants $a,b$.   In the Sen limit, the model reduces to an orientifold of a
Calabi-Yau threefold; one introduces a new coordinate $\xi$, and the equations defining the
Calabi-Yau threefold are (\ref{base}) together with the additional equation
\be
\xi^2 = h(z_i)~.
\ee
The orientifold action then reverses $\xi$ while simultaneously acting with $\Omega ~(-1)^{F_L}$
(where $\Omega$ here denotes worldsheet orientation reversal).

In the Sen construction, there are D7-branes and O7-planes wrapping divisors in the
Calabi-Yau threefold.  The D7 branes
are located on the loci $\eta = 0$, while the O7 planes wrap $h=0$.\footnote{ Strictly speaking, at small
but finite $C$, the O7 planes split into various $(p,q)$ D7-branes, but this is an exponentially
small effect at weak coupling.}

With our choices above,  $\eta$ and $h$ are invariant under the group ${\bf P}$ of all permutations of the four $z_i$'s, $i=1\cdots 4$. This is a discrete subgroup of $O(4)$. There is also a
symmetry of the fourfold (\ref{fiber}) under which
 $z_a\rightarrow e^{2\pi i \over 4} z_a$ for $a = 1,2,3,4$, $z_{5} \rightarrow z_{5}$, and
 $z_{6} \rightarrow - z_{6}$,
which is a ${\mathbb Z}_{4} \subset U(1)_R$.\footnote{The defining equation of the base $B$ (\ref{base}) is
not invariant for non-zero $\epsilon$.  However, by choosing appropriate fluxes, we can
stabilize $\epsilon$ very close to zero \cite{GKP}; this breaking of ${\mathbb Z}_{4}$
can then be made to occur naturally at a tunably small scale, and will not concern us.}

Finally, the action of ${\mathbb Z}_{2 {\rm exchange}}$ in the IIB theory involves a  permutation of  the $z_i$'s combined with $\Omega (-1)^{F_L}$. This too is a symmetry.
We will see in \S5.2\ that this discrete symmetry group is big enough to
disallow any GSROs.
More generally the same discrete symmetries are preserved if $\eta$ is a more general
quartic polynomial invariant under the
permutation group, ${\bf P}$,  containing only  monomials that preserve the ${\mathbb Z}_{4} \subset U(1)_R$ mentioned above.

To complete the discussion, let us calculate the Euler number $\chi$ of our fourfold. It is relevant in determining the length of our throat, and hence the size of the hierarchy obtainable in our compact model.
This is because tadpole conditions bound the allowed three-form flux in the orientifold limit (or
more generally, the four-form flux in the fourfold compactification) to satisfy:
\begin{equation}
KM \leq {\chi \over 24}
\end{equation}
where $K$ is the number of KS cascade steps, and $M$ is the number of fractional D5-branes.
The hierarchy of energy scales generated for a given choice of $K$ and $M$ is of the order
$e^{-{2\pi K\over 3 g_sM}}$ \cite{GKP}.

Using a result in \cite{guk_witt} we have:

\be
\frac{\chi}{24} = 12 + 15 \displaystyle{\int_B} ~ c_1(B)^3 = 492~,
\ee
where $c_1(B)$ is the first Chern class of our 3-fold base $B$. In our example, $c_1(B) = 2J$ where $J$ is the restriction of the K\"ahler form from the ambient ${P}^5$,  and $\int_B J^3 = 4$.
The above result follows. This is a healthy Euler number for our purposes, easily accomodating large
enough fluxes to generate a sizeable hierarchy (and to additionally stabilize $\epsilon$ at a very
small scale).

We can now orbifold this theory by our ${\mathbb Z}_{2}$ operation of $\S4$, $z_{1,2,3,4} \to - z_{1,2,3,4}$, with
$(z_5,  z_6, x, y, z) \to (z_5, z_6, x, y, z)$.  This symmetry acts freely on the geometry, and acts as in \S4\ on the conifold
throat. A ${\mathbb Z}_{2} \subset {\mathbb Z}_{4}$ of the R-symmetry that was preserved acts non-trivially on the orbifold.
Hence the global discrete symmetries preserved  after the orbifolding consist of the permutation group ${\bf P}$,   ${\mathbb Z}_{2 {\rm exchange}}$ and
${\mathbb Z}_{2}$.

\subsection{Discrete Symmetries and GSROs}
In this subsection we show that the discrete symmetries which survive in the compact example constructed above are enough to prevent any relevant operators from arising in the
warped throat region.

A detailed analysis has already been carried out in the non-susy quiver gauge theory above using its  $SU(2)\times SU(2) \times {\mathbb Z}_{M}\times {\mathbb Z}_{2 {\rm exchange}}$ global symmetry(there is an additional $U(1)_B$ symmetry but it does not serve a useful purpose in preventing  GSROs, and we will not include it in the discussion below).
We remind the reader that the ${\mathbb Z}_{M}$ discrete symmetry is a subgroup of the R-symmetry group in the parent susy theory and arises as follows. The KW theory has a $U(1)_R$ symmetry, this is broken to ${\mathbb Z}_{2M}$ in the KS case by the  three form flux. In the  non-susy orbifold a ${\mathbb Z}_{M}\in {\mathbb Z}_{2M}$ acts non-trivially and is a global symmetry.
We saw in the previous subsection that in the compact case a ${\mathbb Z}_{4}$ subgroup
of $U(1)_R$ is left unbroken. For $M\ge 2$ this \footnote{In fact to obtain the orbifold we need $M=2m$, with $m>1$ so this is no restriction.} ${\mathbb Z}_{4} \in {\mathbb Z}_{2M}$.
And a ${\mathbb Z}_2$ subgroup of this $ {\mathbb Z}_{4}$  then acts non-trivially in the orbifold theory.
We denote this ${\mathbb Z}_2$ symmetry as ${\mathbb Z}_{2R}$ below.
In the compact case  we see then that the global symmetries which survive are ${\bf P}, {\mathbb Z}_{2 {\rm exchange}}$ and $\mathbb{Z}_{2R}$.

We now examine whether any GSROs are allowed by this discrete group.
Let us first consider  operators which are singlets under $SU(2)\times SU(2)$,
and ask whether they are ruled out by the  symmetries ${\mathbb Z}_{2 {\rm exchange}} \times {\mathbb Z}_{2R}$.
Since  $Tr(|A|^2-|B|^2)$ is odd under ${\mathbb Z}_{2 {\rm exchange}}$ its descendants in the orbifold theory are not GSROs. At dimension $3$ there are three operators, the two gauginos bilinears and the scalar quartic, $Tr(ABAB)$ in the KW/KS theory. All of  these have R-charge $2$ under the $U(1)_R$, this means they are odd under ${\mathbb Z}_{2R}$  and thus transform non-trivially under it.\footnote{The reader should not
be confused by the notation.  The ${\mathbb Z}_{2R}$ generator acts on the $A,B$ and gaugino fields the
same way that the ${\mathbb Z}_{4}$ generator did in the parent theory, so the gaugino bilinear is ${\it odd}$
under the ${\mathbb Z}_{2R}$.  The symmetry is reduced to a ${\mathbb Z}_2$ because the square of the generator
relates field configurations that our orbifold action has already identified.}
It then follows that the operators which descend from these dimension $3$ operators in the orbifold theory are also not global singlets.
Finally, as we discussed at some length in \S4,  there are no double trace operators which are relevant (of dimension $<4$) in the orbifold theory.

Next, consider operators which are not invariant   under the continuous  $SU(2)\times SU(2)$ group, but which could be invariant under the surviving discrete symmetries. There are essentially three candidates:

 1) First, the KW theory has the operator $Tr(AB)$ which is a $(1/2,1/2)$ under $SU(2)\times SU(2)$. However,
as was discussed in \S4, this operator is not invariant under the ${\mathbb Z}_2$ orbifold symmetry and as a result there are no gauge invariant operators in the orbifold theory which descend from it and which are bilinear in the scalars.

2) Next, there are dimension $2$ operators  in the KW theory which  are the partners (under the superconformal symmetries) of the two $SU(2)$ currents.
These transform like $(1,0)$ and $(0,1)$ representations of $SU(2)\times SU(2)$.
The permutation symmetry $\bf{P}$,  includes three elements which are rotations by
$\pi$ along the three axes of the first $SU(2)$ and also  three   elements which are rotations by $\pi$ along the three axes for the
second $SU(2)$. This is discussed in greater detail in appendix A.
Any operator which lies in the $(1,0)$ representation must transform under
the rotations by $\pi$  along the three axes of the first $SU(2)$ and  cannot be invariant under the permutation symmetry. Thus it cannot give rise to any GSROs in the orbifold theory.
Similarly no GSROs can arise  from the operator which transforms in the $(0,1)$ representation.

3)  This leaves only one other possibility. There is a non-chiral operator in the
KW theory with dimension $3.29$. It arises from vector multiplet I in the classification used in \cite{Ceresole}, see also \cite{AAB}.
This operator transforms as a $(1,1)$ representation  under $SU(2)\times SU(2)$. It is schematically of the form, $Tr(AB \bar{A}\bar{B})$ where the indices are contracted in a gauge invariant manner, and thus is $R$-charge neutral and also neutral under ${\mathbb Z}_{2 {\rm exchange}}$. However, once again, any
  element of the $(1,1)$ representation must transform under the six rotations by $\pi$ mentioned above and thus cannot be a singlet. Therefore, no descendent of this operators in the orbifold theory can give rise to a   GSRO either.

It is difficult to think of any   other operators in the KW theory from which  GSRO descendants might arise in the orbifold theory.  However, to be certain, we have worked through the list of operators in \cite{Ceresole}, applying the selection rules which govern the choice of the $R$ charge (specified by $r$), for given $SU(2)\times SU(2)$ quantum numbers (specified by $(j,l)$). We find that indeed no other  GSROs arise in the orbifold theory. The essential point is that the anomalous dimension grows rapidly with increasing $j,l$, thus beyond  modest values of these quantum numbers no worrisome candidates can arise. We will spare the reader further  details.

\section{Adding interesting physics in the IR}

While our focus in this paper has been to exhibit a large class of field theories without GSROs, at the next step in any
program for realizing the basic picture of \cite{RS}, one would like to find ways to add interesting gauge and matter
sectors in the IR (or, perhaps, spread across the 5th dimension).  Here, we discuss in a very preliminary way two natural
methods of adding interesting matter sectors to our example of \S4, \S5.   While neither gives rise to anything resembling the
Standard Model in detail, both methods illustrate how one may achieve the addition of matter and gauge fields to these
backgrounds without destabilizing the hierarchy.

\subsection{Adding bulk gauge fields: D7s in the throat}

One standard way of adding bulk matter fields in AdS/CFT is to add probe D7-branes to the throat
geometry.  In fact, in the Sen limit of an F-theory compactification, one automatically has an O7-plane
stretched along the locus $h = 0$ and a pair of coincident D7-branes along the locus
$\eta = 0$ \cite{Sen}, where for our concrete model the polynomials $h$ and $\eta$ are given in
(\ref{defh}) and (\ref{defeta}).

It is clear from (\ref{base}) that the deformed conifold singularity of interest to us in the orientifold arises in the patch
$z_5 = 1$ with small values of the $z_i$ satisfying
\be
\sum_{i=1}^{4} z_i^2  + {\cal O} (\epsilon^2 z_i^4)= t^2~.
\ee
Therefore, for suitable choices of parameters, we can arrange for the D7-branes in the geometry
to stretch into the throat region (while, for simplicity, leaving the O7-plane far away).

Concretely, taking a small value of $a$ in (\ref{defeta}), we see that the zero locus of $\eta$ will pass arbitrarily close to the deformed tip of the
conifold, while the O7-plane remains localized far away for $b$ of ${\cal O}(1)$.  In this limit, we obtain
a bulk $U(2)$ gauge theory from the D7-branes stretching down the conifold throat.

The ${\mathbb Z}_2$ orbifolding which breaks SUSY acts freely on the surface $\eta = 0$ wrapped by the
D7s, so it simply changes the topology of the divisor which the D7-branes wrap.   The moduli of
the D7-branes are geometrized in F-theory as deformations of the fourfold complex structure.
Assuming the fourfold complex structure moduli are stabilized by fluxes at a high scale,
as in \cite{KKLT}, the low-energy theory on the D7-branes will then be a pure $U(2)$ gauge theory.

In this way of adding bulk gauge fields to the throat, there is no danger of destabilizing the
hierarchy.  The symmetries of the geometry already eliminated any GSROs, and the D7 embedding
arises here for specific choices of the symmetric geometry.
On the other hand, the small value of $a$ we require to obtain D7s which live far down the throat
is not explained at this level;  it is a tune that needs to be attributed to the details of moduli
stabilization.
It is natural to ask if there are simple mechanisms that would guarantee the stabilization of the D7s with exponentially small $a$ (i.e. stretching far down the throat); we expect such mechanisms can be found, but leave this
for future work.

\subsection{Higgsing a group in the deep IR: Anti D3-branes in the throat}

Another natural ingredient in the models of this class is D3-branes.  However, in the relevant tree-level
solutions of IIB supergravity, D3-branes feel no force.  Therefore, any probe D3-branes may be driven out of the throat
by sub-leading corrections; they may suffer an instability to run away on their Coulomb branch.
While this is presumably model dependent, it would require further work to analyze under what
circumstances probe D3s would be stable in the IR region of the throat.  This is the reason we
focused on the case $N=kM$ in the discussion of \S4; then the cascade ends with no left-over
probes.

Instead, we can add probe anti-D3s.  As in \cite{KPV}, they will be pulled to the tip of the
(now orbifolded) KS geometry by the background 5-form flux.   If we add $p$ such anti-D3s
(with $p << M, N$ to retain calculational control), then their fate is the following: the $SU(p)$
gauge theory on the anti-D3s is Higgsed at an exponentially low-scale (by the anti-D3 adjoint
scalar fields) in a way that completely breaks the gauge symmetry.  This is seen via a Myers
effect in the flux background at the tip of the geometry \cite{KPV}.  The interpretation of
these objects as states in the dual field theory has been discussed in \cite{DKM}.

Unlike the D7s, the probe anti-D3s ${\it do}$ break the global symmetry group in an important way.
For instance, their positions break the $SU(2)^2$ isometries.  However, this spontaneous breaking
of the global symmetries in the IR is not dangerous, for the same reason the breaking of
${\mathbb Z}_{2M} \to {\mathbb Z}_2$ in the KS theory is not dangerous -- it happens in the deep IR, and the possible
subsequent generation of
relevant perturbations to the field theory at such a low scale does not destabilize the
hierarchy.

Therefore, this example gives a concrete instance of a non-Abelian gauge theory undergoing the
Higgs mechanism at energy scales $<< M_{\rm Planck}$ in a theory with high-scale SUSY
breaking.

One rather interesting feature of this example is that the emergent $SU(p)$ gauge theory
can be weakly coupled (although it emerged from the cascading strongly coupled large $N$
gauge theory).  In addition, there are no $SU(p)$ charged bulk gravity modes: only the
open-string states stretching between the anti-D3s carry $SU(p)$ gauge quantum numbers.
The excited string states are parametrically heavier than the KK modes at the end of the throat.
This also implies that $ p << M, N $ should be the relevant species factor controlling
radiative corrections to the anti-D3 gauge theory.  This could be important in obtaining
reasonable values of precision electroweak observables such as the $S$ and $T$ parameters.

Furthermore, because of the nature of the cascading gauge theory, the effective 't Hooft coupling
of the hidden approximate CFT is considerably smaller in the
IR than the coupling in the UV ($g_s M << g_s N$).
This means that one may be able to use supergravity to control the hierarchy, while just
approaching the border of (or even leaving) the supergravity regime in the IR region.
This has been observed to improve, e.g., the nature of the electroweak phase transition
in RS models \cite{JMR}.    So we see that the presence of several distinct expansions
(as opposed to a single 't Hooft expansion in $g_s N$) in the more detailed string constructions
offers some qualitative hope of solving the phenomenological problems of the
simplest large-$N$ toy models.

It would be interesting to try and generalize the work of \cite{Quevedo} to give more realistic
models in these completely non-supersymmetric throats.


\bigskip
\centerline{\bf{Acknowledgements}}

We are grateful to K. Agashe, A. Cohen, A. Dymarsky, J. March-Russell, L. McAllister, M. Strassler, and R. Sundrum for extensive discussions
about related subjects (over a period of years).  We also thank D. Morrison for serving as a cheerful consultant about annoying mathematical details, and J. Polchinski, R. Sundrum and S. Yaida for providing thoughtful comments on a draft.   This work was supported by the Stanford
Institute for Theoretical Physics, the NSF under grant PHY-0244728, and the DOE under
contract DE-AC03-76SF00515.  D.S. is supported by the Mayfield
Stanford Graduate Fellowship.  S.K. and D.S. acknowledge the kind hospitality of the Kavli Institute for
Theoretical Physics during the completion of this work, and the support of NSF grant PHY05-51164. S.P.T. is on a sabbatical visit  to Stanford University and SLAC National Accelerator Laboratory for the period 2008-2009. He  thanks his hosts for their  support and kind hospitality.  He also acknowledges support from the Swarnajayanti Fellowship, Govt. of India, and thanks the organisers and participants of  the two workshops, ``From Strings to LHC I, II'', held at Goa and Bangalore respectively, for some very helpful discussions.
Most of all S.P.T. would like to thank   the  people of India for generously supporting  research in String Theory.

\bigskip

\appendix
\section{The action of ${\bf P}$ on different $SU(2) \times SU(2)$ representations
}
The permutation group, ${\bf P}$ was introduced in our discussion of the discrete symmetries preserved by the compact Calabi-Yau manifold, in \S5.2.
Here we discuss how this group is embedded in the continuous group $SU(2)\times SU(2)$. This will allow us to determine how ${\bf P}$  acts on any representation of $SU(2)\times SU(2)$.

 The four coordinates, $z^i$, in eq.(\ref{fiber}), eq.(\ref{base}), transform as a $(2,2)$ representation
 of $SU(2)\times SU(2)$. This means on  the matrix,
\be
\label{actionm}
 M=\pmatrix{ z_1+iz_2 & z_3+iz_4 \cr
 -(z_3-iz_4) & z_1-iz_2 \cr},
\ee
we can take  the first $SU(2)$ to act on the left, $M\rightarrow U\cdot M, U\in SU(2)$,
and  the second $SU(2)$ to act on the right similarly.

Six elements of ${\bf P}$ in particular played an important role in our discussion of the GSROs above. These are the three
rotations by angle $\pi$ about the three axes of the first $SU(2)$s, and similarly the second $SU(2)$.  Consider  a rotation by $\pi$ about the $z$-axis of the first
$SU(2)$. It  acts on $M$ by the matrix
\be
\label{defu1}
U=\pmatrix{e^{i\pi\over2} & 0 \cr 0 &  e^{-i\pi\over 2}\cr}
\ee
acting on the left.
Under it, $(z_1,z_2) \rightarrow (-z_2,z_1)$ and
$(z_3,z_4)\rightarrow (-z_4,z_3)$.
It is easy to see that this keeps the polynomials $\eta$ and $h$, eq.(\ref{defeta}), eq.(\ref{defh}),
invariant and  is therefore  a symmetry of the Calabi-Yau.
Similarly for all the other rotations by angle $\pi$.

To be more thorough, the group ${\bf P}$ consists of $6$  pair-wise exchanges and elements of order $3$ and $4$ obtained by composing these pair-wise exchanges.
Now an  exchange of any two coordinates, say $z^1$ and $z^2$, is carried out by the matrix
\be
\label{matex}
\pmatrix{z^1 \cr z^2\cr}\rightarrow \pmatrix{0&1\cr1&0\cr}\pmatrix{z^1\cr z^2\cr}
\ee
which has determinant $-1$. This lies in $O(4)$ but not in $SU(2)\times SU(2)$.
A related symmetry which does lie in $SU(2)\times SU(2)$, is obtained by composing the exchange above with an inversion, in this case say, $z^1\rightarrow -z^1$, (with the other coordinates held fixed). The resulting transformation is now carried out by the matrix $\pmatrix{0&-1\cr1&0\cr}$,
 with determinant $+1$. Since the inversion is also a symmetry of the Calabi Yau manifold, this final transformation is also a symmetry.

In this way we can obtain pair-wise exchange elements (by appending  additional signs) which are all elements of $SU(2)\times SU(2)$. The order $3$ and $4$ elements obtained by further composing them are then automatically also elements of $SU(2)\times SU(2)$. We take ${\bf P}$ to be the resulting group of permutations obtained in this manner.
By construction it is a now a subgroup of $SU(2)\times SU(2)$. And in particular the rotations by angle $\pi$ about the axes of the first and second $SU(2)$s are then all elements of ${\bf P}$. More generally, it is easy to determine how the pair-wise exchanges act on any representation of $SU(2)\times SU(2)$, and from there find how all elements of ${\bf P}$ act on the representation.

\section{The Horowitz-Orgera-Polchinski instability}

The Horowitz-Orgera-Polchinski (HOP) instability of $AdS_5 \times S^5/{\mathbb Z}_{k}$ compactifications \cite{joegary},
with the ${\mathbb Z}_{k}$ action given by (\ref{genzn}) with
$k>3$ and odd, can be understood as follows.  $S^5$ can be viewed as a circle fibration over
$CP^2$, with metric
\be
ds^2 = R^2 \left( ds_{CP^2}^2 + (d\chi + A)^2 \right)
\ee
where $\chi$ is the coordinate on the circle fiber and $A$ is a gauge connection (of the
KK gauge field) on $CP^2$.
If the periodicity of $\chi$ is taken to be $2\pi$ on the original $S^5$, then on $S^5/{\mathbb Z}_{k}$, the
periodicity becomes $2\pi/k$.  The orbifolded circle then has circumference $2\pi R/k$.

The vacuum energy of a string stretching around this orbifolded circle is given by:
\be
\alpha^\prime M^2 = {R^2 \over {\alpha^\prime k^2}} + {2(3-k) \over k}
\ee
which is always positive at large 't Hooft coupling, but becomes tachyonic at weak
't Hooft coupling (small $R$).

The boundary conditions on fermions encircling the minimal circle in the quotient, due to the
factor of $(-1)^F$ in (\ref{genzn}), are anti-periodic.
While the tachyonic instability described above for small $R$ is a stringy effect,
Witten demonstrated long ago that the Kaluza-Klein vacuum $R^4 \times S^1$  on a circle with anti-periodic boundary conditions for fermions is unstable even in the large
radius limit; there is a tunneling instability induced by a ``bubble of nothing" \cite{Witten}.
In the Euclidean solution describing false vacuum decay, the $S^1$ shrinks smoothly to a point as one moves
in from infinity in the radial direction of $R^4$; the solution is in fact the analytic continuation
of the Schwarzschild solution.

The main insight of the HOP paper is that a similar bubble of nothing solution exists for the
$AdS_5 \times S^5/{\mathbb Z}_{k}$ orbifolds under consideration.  The intuition is that the $S^1$ fiber over
the $CP^2$ in $S^5$ plays the role of the $S^1$ in Witten's analysis.
A fascinating new ingredient is that since in a conformal theory there is no scale, the decay rate
must be either 0 or infinite; the HOP analysis shows that the integral over the value of the
radial coordinate where the bubble nucleates gives an infinite rate.  The non-compact
$AdS_5 \times S^5/{\mathbb Z}_{k}$ orbifold thus decays instantly.\footnote{More precisely, a bubble nucleates
${\it somewhere}$ instantly; a given observer will be struck in about an AdS Hubble time.}

This sounds like it would have dramatic effects for our discussion, but in fact it does not.
If one were to take the ${\cal N}=4$ orbifolds of \S2, and couple them to 4d gravity as in \S5,
the decay rate computed in \cite{joegary} is regulated by the UV cutoff.  Instead of integrating
the decay rate over the entire radial direction of $AdS_5$, the integral is cut off at some finite
$r_{UV}$.  The rate then becomes negligibly small.  It is estimated in equation (5.2) of
\cite{joegary}; the result is that with a cutoff at energy scale $\Lambda$, one finds an integrated
rate
\be
\label{rateis}
\Gamma \sim k^9 e^{-B} \Lambda^4
\ee
where the instanton action is
\be
B \sim N^2/k^8~.
\ee
In the limit of large $N$ with fixed $k$, this vanishes rapidly.
Our conclusion is that cut-off throats based on the orbifolds of \S2\ are viable despite the
existence of the HOP instability; their lifetimes can be made cosmologically realistic.

The case of the cascading theory of \S4\ is more involved.  As described in \cite{joegary}, the
growth of the effective number of colors $N$ with the radial coordinate in cascading theories
renders their integral of the decay rate over the radial direction ${\it finite}$, even in the
non-compact cascading solution.  We conclude that for our construction involving quotients of
the Klebanov-Strassler throat, the HOP instability, while again an interesting feature, does not
provide a serious constraint.

\end{document}